# Computational modeling approaches in gonadotropin signaling

**Mohammed Akli Ayoub, Romain Yvinec, Pascale Crépieux and Anne Poupon***

BIOS group, INRA**,** UMR85, Unité Physiologie de la Reproduction et des Comportements, F-37380 Nouzilly, France ; CNRS**,** UMR7247, F-37380 Nouzilly, France ; Université François Rabelais, F-37041 Tours, France

***Corresponding author:**

Anne Poupon, UMR PRC, F-37380 Nouzilly, France.
Phone : + 33 2 47 42 75 05 ; Fax : + 33 2 47 42 77 43 ; e-mail : Anne.Poupon@tours.inra.fr

**Abstract**

Follicle-stimulating hormone (FSH) and luteinizing hormone (LH) play essential roles in animal reproduction. They exert their function through binding to their cognate receptors, which belong to the large family of G protein-coupled receptors (GPCRs). This recognition at the plasma membrane triggers a plethora of cellular events, whose processing and integration ultimately lead to an adapted biological response. Understanding the nature and the kinetics of these events is essential for innovative approaches in drug discovery. The study and manipulation of such complex systems requires the use of computational modeling approaches combined with robust *in vitro* functional assays for calibration and validation. Modeling brings a detailed understanding of the system and can also be used to understand why existing drugs do not work as well as expected, and how to design more efficient ones.

**Introduction**

Follicle-stimulating hormone and luteinizing hormone receptors (FSHR and LHR) play central roles in animal reproduction. These two receptors belonging to the family of G protein-coupled receptors (GPCRs) are able to transduce the signals mediated by the variations of FSH and LH blood concentrations in adapted cellular responses in gonads, mostly through the control of follicle development in the ovary and gametogenesis in the testis (Ascoli et al., 2002; Simoni et al., 1997). Binding of the hormone to its cognate receptor triggers cascades of biochemical reactions within the cell, which results in profound changes in gene transcription (Friedmann et al., 2005) and protein translation (León et al., 2014; Musnier et al., 2012). These signaling cascades mainly originate from the interaction of the activated receptors with G proteins on the one hand, and ß-arrestins on the other hand (Gloaguen et al., 2011; Landomiel et al., 2014; Ulloa-Aguirre et al., 2011). The precise balance between these two pathways is essential for the biological outcome of cell stimulation. As an example, the modification of this balance in the N680S single nucleotide polymorphism in the FSHR is thought to be responsible for the relative resistance of S/S women to FSH treatments (Tranchant et al., 2012).

GPCRs being membrane receptors have a high specificity for their ligands, and are often individually expressed in a reduced number of cell types. Moreover, they are involved in most cellular processes. Consequently, they are ideal drug targets being the target of about 40% of marketed drugs (Ma and Zemmel, 2002). Some of these drugs are balanced full agonists, meaning that they trigger the same signaling pathways than the natural agonists, with the same balance between G-dependent and ß-arrestin-dependent pathways. During the last decade, it has been shown that it was possible to find GPCR ligands, generally referred to as "biased", which trigger only part of the natural ligand's signaling repertoire. From the therapeutic point of view, such pharmacological profile are thought to be more efficient and safer since only the beneficial pathways (G proteins or ß-arrestins) are targeted. The most famous example of such use of GPCR biased ligand is carvedilol (Wisler et al., 2007), one of the ß-blockers used in heart failure pathologies. By blocking the ß-adrenergic receptor, they reduce the catecholamine stimulation in heart and other organs. Contrarily to most ß-blockers which are full antagonists, carvedilol only blocks the G-dependent pathways, leading to therapeutic efficacy comparable to the one of other ß-blockers, but it extends survival (Poole-Wilson et al., 2003).

Although the events following ligand binding are referred to as signaling pathways, the term of signaling networks better represents the complexity of these events and the tight interconnections between them. This complexity can be envisaged at different levels. (1) A large number of molecules participate to this network, and there are many regulation feedback loops within it. (2) The existence of homo- and hetero-dimers has been demonstrated for many GPCRs (Parmentier, 2015) (je citerai ma revue dans Annu Rev Pharm Toxico). (3) The biological outcome depends on the duration of the stimulation and on the stimulation profile (short *vs* long, continuous *vs* pulsatile). (4) *In vivo* the cell is submitted to multiple and coordinated stimulations, and the resulting signaling networks are interlocked. Those points are nicely illustrated by the strong interconnections between LH and FSH, which (1) share large portions of intracellular mechanisms triggered by their cognate receptors (Figure 1), (2) whose receptors have been shown to heterodimerize (Mazurkiewicz et al., 2015), (3) have very precise and coordinated profiles,-and antagonistic cellular effects, (4) and of course both are involved in the control of reproduction.

The pharmacology of gonadotropin receptors is still very poor as only natural hormones are used, either extractive in animal reproduction, or recombinant in human health. These last years have seen the appearance of many small molecules (Nataraja et al., 2015), either targeting the orthosteric site (as the natural ligand), or allosteric modulators (modulating the action of the natural ligand) of the FSHR and LHR (+ Dias.J et al, MCE 2011). However, optimizing the action of such pharmacological molecules requires the detailed knowledge of their action, which in turn depends on the precise knowledge of the signaling networks. Over the last decade it has became increasingly evident that computational modeling is a tool of choice for such purpose.

**Modeling the signaling network activation**

Signaling networks trigger complex cellular responses such as proliferation, differentiation, apoptosis, through the regulation of transcription and translation processes. Each of these responses results from precise kinetic activation/deactivation profiles of the molecules constituting the network. Therefore, it is of paramount importance to understand the major molecular mechanisms set off by the binding of a given ligand to the receptor, how this signal propagates throughout the signaling network, and how the different profiles are integrated in adapted cellular responses. The complexity of these networks in itself prohibits the direct measurement of all the kinetic profiles. Moreover, for many of these molecules, the tools are not available. Building computational kinetic models of the network allows accessing these profiles by measuring only a small subset or readouts.

Two main types of models should be distinguished: topological (or static) models, which list the molecules belonging to the network and the relations between them (activation, inhibition, phosphorylation, etc.); and dynamic models, which describe the precise kinetic evolution of the molecule concentrations.

Topological models are a necessary first step in any modeling process. Indeed, building the model entails a detailed study of the literature, and addresses questions that would otherwise remain unresolved. This model can also be used to answer qualitatively to the question "can my data be obtained with this model?". At first try the answer is usually negative, and iterations between modeling and experimental results are always necessary before a satisfying model is found. We have done this work for the FSH model (Gloaguen et al., 2011), and it involved the analysis of more than 150 papers, from which more than 200 experiments were extracted and used for building the model. Figure 1 shows a schematic view of this model.

GPCR-induced signaling events involve fast and reversible reactions. Consequently, static models cannot faithfully represent the outcome of receptor activation. Moreover, the time-scale of the different pathways is often different, and some molecules have a clear biphasic profile. The best known example is the bimodal phosphorylation of ERK, which occurs in seconds through the G-dependent pathway, but lasts only a few minutes, then after 10 minutes depends on the ß-arrestin dependent pathway, and extends beyond an hour. Static models cannot represent such biphasic behavior. Even worse is the case of cAMP since the production of cAMP is activated by the Gαs pathway, but inhibited by the ß-arrestin pathway upon the activation of a given receptor.

It is therefore fundamentally important to take dynamics into consideration when trying to model and predict the outcome of receptor activation. The ordinary differential equations (ODE) formalism is the most popular for dynamic models. In this formalism, the concentration variation of a given molecule is expressed as a function of time, and depends on concentrations of the other molecules within the network. Different rate laws for the reactions within the network can be chosen. The simplest kinetic law is derived from the law of mass action (Figure 2A and B). This model is built from the static model, adding kinetic constants on the reactions and initial concentrations of the different molecules.

**Estimating parameters to predict biological responses**

The major problem at this point is that these constants and concentration are usually unknown, and for most of them not accessible. However, computational methods exist to estimate these parameters from a limited number of experimental data. The principle of parameter estimation is to search for sets of values that, when used to simulate the model dynamics, reproduce experimental data (Figure 2C and D). We have developed such a method, and applied it to the modeling of the angiontensin signaling pathway (Heitzler et al., 2012). We have shown that when the simulation tightly fits experimental data, it is possible to predict the behavior of the system in biological conditions different from those used to acquire initial data. For example, the opposite effects of ß-arrestins 1 and 2 on the phosphorylation of ERK (by inhibiting or activating, respectively) is correctly predicted. Moreover, the predicted dose-responses of ERK phosphorylation in control conditions and when ß-arrestin 1 or ß-arrestin 2 are depleted are very close to the measured ones (Figure 3).

From this model we were able to propose hypotheses that were then validated experimentally. For example, the model predicted that the desensitization pathway exerts a strong inhibition on the ß-arrestin-dependent signaling pathway. This hypothesis has been validated in HEK293 cells, but also in primary vascular smooth muscle cells, which naturally express the angiotensin II (AT1) receptor. We were also able to show that this mechanism exists for other GPCRs, including the FSHR.

**Obtaining experimental data to build models**

Because the calibration of the model is made through fitting experimental data, the quality of the final model is tightly linked to the quality of these data. Until recently, the method of choice for following the kinetics of signaling events has been the measurement of phosphorylation in Western blot assays at different time points. However, this method has many drawbacks in terms of computational modeling. First, the different time points correspond to different cell cultures. Even if the biological conditions are very tightly controlled, there will always be some minor differences between the different wells. Second, it cannot be envisaged to acquire data at very close time points and on a long period. Consequently, there are multiple solutions for drawing a curve that passes through all the experimental point (Figure 4A).

In that regard, newly developed methods based on fluorescence and luminescence energy transfer allowing measuring cellular events in living cells are very precious. Indeed, they allow to acquire the complete kinetic profile on the same cells, and to have a very short interval between time points, guarantying uniqueness of the solution (Figure 4B).

Such assays have been developed and were recently applied to the study of FSHR and LHR signaling networks (Ayoub et al., 2015). Using bioluminescence and fluorescence resonance energy transfer (BRET and FRET), we were able to measure the activation of heterotrimeric G protein, production of cAMP, calcium release, ß-arrestin 2 recruitment and receptor internalization and recycling in real-time and live cells (Figure 5). By measuring kinetic profiles at different hormone concentrations, it is also possible to obtain dose-response curves for each pathway (Figure 6). Such an analysis led to interesting observations in terms of how can the different pathways triggered by a given recepotor be regulated. Indeed, it appears that there is a 3 log difference in sensitivity to ligand concentration between the G-dependent and ß-arrestin-dependent pathways, which is very unusual for GPCRs. Now remains the task of understanding the role of this difference in the physiology of gonadal cells.

**Affinity, efficacy and potency**

Receptor-ligand interactions are classically characterized by two essential parameters: i) affinity, which reflects the energy of the ligand-receptor interaction and ii) intrinsic efficacy, which measures a biological response to ligand binding. The combination of these two parameters defines the potency (Figure 7A). Consequently, two ligands having different affinities can exhibit the same potency if the less affine ligand has better intrinsic efficacy. This point is very important in the context of drug screening, since the affinity is not necessarily the best indicator of a good lead.

However, in the case of GPCRs, and probably for other types of membrane receptors, this model is an oversimplification. Indeed, the classical determination of intrinsic efficacy relies on the hypothesis of a single, linear signaling pathway going from the receptor to the measured biological outcome (Figure 7A). However, the efficacy should take into account at least G-dependent and ß-arrestin-dependent signaling (Figure 7B). These two pathways are the most proximal events in the signaling network and difficult to measure. Therefore, more distant outcomes are often chosen, such as cAMP production for the Gs-dependent pathway and ERK and PDE for ß-arrestin which also affect cAMP levels. Consequently, cAMP levels do not accurately reflects Gαs activation, and efficacy and potency cannot be derived straightforwardly from it (Figure 7C).

Consequently, ligand bias, although its mere existence is not questioned any longer, is difficult to evaluate. The idea is to be able to compare the efficacies related to two different effectors, for different ligands. For example, in the case of the FSHR, a ligand biased towards the ß-arrestin pathway would be a ligand eliciting more ß-arrestin activation, relative to Gαs, than the natural ligand (Figure 7D). Different methods have been proposed for computing this bias (Kenakin, 2014; Landomiel et al., 2014; Rajagopal et al., 2011). All these methods have in common to be derived from Black & Leff's operational model (Black and Leff, 1983), which gives a simple mass-action based model to fit  the measurement of dose-responses. As already mentioned for efficacy and potency, in practice this requires the measurement of effectors immediately downstream the receptor at different ligand concentrations. When more distant downstream read-outs are chosen, the distortions introduced in the kinetics of activation by the presence of intermediaries and/or cross-talks renders the calculation mathematically challenging.

By allowing simulating kinetic profiles that are not easily accessible to experimentation, computational modeling can help solve this question. For example, in the angiotensin II signaling network published in (Heitzler et al., 2012), although only DAG, active PKC and phosphorylated ERK were measured experimentally, G activation and ß-arrestin recruitment could be simulated (Figure 8). We were able to partially validate these simulations by the computation of activation half-lives (3.3s for G and 80s for FSHR-ß-arrestin 2) and the comparison with measured values in independent publications (0.3-2s for G (Lohse et al.,

2008; Vilardaga, 2010) and 20-50s for FSHR-ß-arrestin 2 (Lohse et al., 2008; Rajagopal et al., 2006)).

## Conclusions

Methods now exist to model signaling networks. The quality of these models tightly depends on the quality of the experimental data used to calibrate them. In that regard, luminescence and fluorescence-based assays provide irreplaceable constraints for the estimation of unknown parameters. As shown through the angiotensin receptor model, such models allow predicting the behavior of the system in biological conditions different from those used to acquire calibration data. It is thus possible to choose the most interesting conditions, and implement high added-value experiments. Models can also be used to better understand the effects of biased ligands, and consequently pave the way to a better method for estimating biases, but also for understanding how they affect the biological outcome they induce, thereby participating in the development of better drugs, more efficient and safer.

## Figure legends

**Figure 1:** Schematic view of the FHSR and LHR signaling networks.

**Figure 2:** (A) For any chemical reaction, the variation d[B] of the concentration of B in the short time dt can be written as a function of the concentrations of A and B and the rate constants $k_0$ and $k_1$ (equation 1). In the case of the mass action law, this function takes the form indicated in equation 2. (B) When the reaction is catalyzed by molecule C, this function takes the form indicated in equation 3. (C) The dynamic network is built from the static network by adding kinetic rates on reactions and considering the initial molecules concentrations. (D) The error between the experimental values and a simulation, obtained using a given set of parameter values, is a function of the differences between measured and simulated points. (E) To estimate the values of the unknown parameters an iterative method is used at each step, on or more sets of parameter values are simulated, and the corresponding error computed. If the error is not null, the values of parameters are modified. This iterative process stops when the error is null or when the maximal number of iterations is reached.

**Figure 3:** Comparison of simulated and measured dose-responses of ERK phosphorylation triggered by the angiotensin receptor in control conditions, and conditions depleted in either ß-arrestin 1 or ß-arrestin 2.

**Figure 4:** Data obtained through classical Western blot analysis (A) do not introduce a lot of constraints to the model, and many different curves passing through all the experimental points can be drawn. On the contrary, with FRET/BRET data (B), the curves going through all the data points are close to each other.

**Figure 5:** Different BRET assays for the study of FSHR and LHR signaling networks. (A) CAMYEL assay for measuring the kinetics of cAMP production. In the basal state the Rluc and GFP fused to Epac are close to each other, allowing an important energy transfer. When the receptor is activated, cAMP is produced and binds to Epac changing the conformation of Epac and increasing the distance between Rluc and GFP which is measured as a BRET decrease. n=3 for FSHR and LHR. (B) Assessment of G activation. At the basal state, the $\alpha$, $\beta$

and γ subunits of the heterotrimeric G-protein are in complex, bringing close together the Rluc fused to Gα and the Venus fused to Gβγ, allowing an important energy transfer. When the receptor is activated and the Gα is activated, thus dissociating from Gβγ, the distance between Rluc and Venus increases and consequently the BRET signal decreases. n=2 for FSHR, n=1 for LHR (C) Measurement of calcium release. In absence of calcium, aequorin is inactive and not able to transfer energy to GFP. When the receptor is activated and calcium is released, aequorin conformation changes and transfers energy to GFP. n=2 for FSHR, n=3 for LHR. (D) Measurement of ß-arrestin 2 recruitment. At the basal state the Rluc fused to the receptor and the yPET fused to ß-arrestin 2 are too distant for energy transfer. Recruitment of the ß-arrestin 2 to the receptor after activation brings them closely together and the BRET signal increases. n=2 for FSHR and LHR.

**Figure 6:** Dose-response analysis on cAMP production and ß-arrestin 2 recruitment for hFSHR (left) and hLHR (right), the individual dose-response data obtained in each BRET assay as indicated were normalized to the maximal signal taken as 100% of receptor-mediated responses.

**Figure 7**: Measuring affinity, efficacy and potency. In the classical paradigm (A), the affinity is measured by the dissociation constant Kd. To evaluate the potency and efficacy, the quantity of effector is measured as a function of ligand concentration.  The potency (Emax) is the maximal response that can be obtained, and efficacy ($EC_{50}$) is the ligand concentration necessary to obtain half of the maximal response. In a more complex case (B), efficacies and potencies relative to different effectors could still be defined. If effectors are not direct targets of the receptor and intermediary molecules exist between them, or if one effector affects the other one, efficacies and potencies cannot be defined.

**Figure 8:** Simulations of the kinetic profiles of active G protein and FSHR-ß-arrestin 2

signaling complex in the angiotensin signaling network.